\documentclass[12pt]{article}

\usepackage{color}
\usepackage{amsmath,amssymb,tensor}
\usepackage{cite}
\usepackage[affil-it]{authblk}
\usepackage[T1]{fontenc}
\usepackage[labelsep=period]{caption}
\usepackage[unicode,linktocpage=true,plainpages=false,pdfpagelabels=false]{hyperref}

\definecolor{linkcolor}{rgb}{0.6,0,0}
\definecolor{citecolor}{rgb}{0,0.6,0}
\definecolor{urlcolor}{rgb}{0,0,0.9}
\hypersetup{colorlinks, linkcolor={linkcolor},citecolor={citecolor}, urlcolor={urlcolor}}
\newcommand{\dd}{\partial}
\newcommand{\de}{\delta}
\newcommand{\m}{\mu}
\newcommand{\n}{\nu}
\newcommand{\ls}{\left(}
\newcommand{\rs}{\right)}
\newcommand{\ka}{\varkappa}
\newcommand{\ga}{\gamma}
\newcommand{\si}{\sigma}
\newcommand{\ff}{\varphi}
\newcommand{\lks}{\left[}
\newcommand{\rks}{\right]}
\newcommand{\ta}{\tau}
\newcommand{\al}{\alpha}
\newcommand{\be}{\beta}
\newcommand{\Ga}{\Gamma}
\newcommand{\po}{{\Pi_{\!\!\bot}}}
\newcommand{\te}{\theta}

\newcommand{\disn}[2]{$$\displaylines{\refstepcounter{equation}%
            \label{#1}\hskip 1em minus 1em #2\hfilneg}$$}
\newcommand{\nom}{\hfil\hskip 1em minus 1em (\theequation)}
\newcommand{\ns}{\hfill\cr\hfill}

\begin{document}

\title{Gravitational energy in the framework\\
of embedding and splitting theories}
\author{D.~A.~Grad\thanks{E-mail: d.grad@spbu.ru}, R.~V.~Ilin\thanks{E-mail: st030779@student.spbu.ru}, S. A. Paston\thanks{E-mail: s.paston@spbu.ru}, A. A. Sheykin\thanks{E-mail: a.sheykin@spbu.ru}}
\affil{Saint Petersburg State University, Saint Petersburg, Russia}

\date{\vskip 15mm}
\maketitle

\begin{abstract}
We study various definitions of the gravitational field energy
based on the usage of isometric embeddings in the
Regge-Teitelboim approach. For the embedding theory we consider
the coordinate translations on the surface as well as the
coordinate translations in the flat bulk. In the latter case the
independent definition of gravitational energy-momentum tensor
appears as a Noether current corresponding to global inner
symmetry. In the field-theoretic form of this approach (splitting
theory) we consider Noether procedure and the alternative method
of energy-momentum tensor defining by varying the action of the
theory with respect to flat bulk metric.
As a result we obtain energy definition in field-theoretic form
of embedding theory which, among the other
features, gives a nontrivial result for the solutions of embedding
theory which are also solutions of Einstein equations.
The question of energy localization is also discussed.

\end{abstract}

\newpage

\section{Introduction}\label{vved}
One of the oldest problems of general relativity, which arose almost simultaneously with the GR itself, is the problem of gravitational energy definition. The first attempts to define the energy of gravitational field were made by Einstein and Grossman in 1913 \cite{Einstein1913}, even before the GR in its final form were formulated. Einstein's investigations, as it is known, attracted Hilbert's attention, and he devoted several papers to the study of the variational principle properties. In 1918 Emmy Noether proved that there is a regular method which allows one to construct conserved quantities for a given Lagrangian theory, if its action is invariant with respect to any continuous symmetry group. It should be stressed that work of Noether was initially aimed at examining the properties of conserved quantities in gravity \cite{vizgin_noether}.

However, it is known that the result of Noether theorem applying to the GR Lagrangian (in any form: either Einstein-Hilbert or first order one) is coordinate-dependent, so the corresponding Noether current turns out to be the energy-momentum pseudotensor (pEMT) rather than usual EMT.  Later it was shown by Tolman \cite{tolman} that the full EMT which included Einstein pseudotensor (the one that corresponds to first order Lagrangian of gravity) can be written as a divergence of some \textit{superpotential}, and Freud obtained \cite{Freud} an antisymmetric form of such a superpotential, from which the vanishing of its divergence is obvious. For the review of various kinds of gravitational pEMTs and superpotentials, see \cite{Petrov-gr-qc/0705.0019}.

The new phase of gravitational energy studying began with the renowned paper of Arnowitt, Deser and Misner \cite{adm}, who constructed the Hamiltonian for 3+1-splitted spacetime. A detailed discussion of the energy problem in Hamiltonian formulation of gravity can be found in L. D. Faddeev's paper \cite{faddeev-UFN1982}. In particular, in this paper he stressed that if one chooses an action in its first order form (as in \cite{Einstein1916}) then the full energy corresponding to that action turns out to be positive and vanishes only in the absence of matter sources and gravitational waves.

In 1975 Regge and Teitelboim, the authors of the prominent paper \cite{RT1974} about the problems of Hamiltonian approach to gravity, proposed a new way to solve these problems. Inspired by successes of string approach, they suggested \cite{regge}  to consider gravity as a dynamics of 4D surface which is locally isometrically embedded in a flat ambient spacetime (bulk). From their point of view the existence of the well-defined time direction in this flat spacetime  could potentially be of use in the canonical quantization of such a theory.

The main purpose of this paper is the investigation of various possibilities of defining conserved energy in the embedding approach proposed by Regge and Teitelboim. Note that the existence of the Minkowski metric in the ambient space gives us a way to construct the gravitational EMT through the usual field theory procedure, namely by varying the action with respect to ambient space metric\footnote{A similar procedure is possible in the usual metric formulation when one has an arbitrary background metric $\eta_{\mu\nu}$: $g_{\mu\nu} = \eta_{\mu\nu}+h_{\mu\nu}$ \cite{Grishchuk1984}. However, in this approach the background problem arises.}. We consider such an approach for the field-theoretic form of the embedding theory.

The section~\ref{en-vlog} begins with a short review of the essential ideas of the embedding approach. After that we perform Noether procedure in two ways: for translations of timelike coordinate on the surface and for translations of ambient Minkowski time.
The relation between the obtained results and GR ones is discussed then.
In the beginning of the section~\ref{en-razb} we shortly describe the form of embedding theory proposed in \cite{statja25} (so-called splitting theory) which has the form of the some field theory in a flat spacetime of high dimension. Then we calculate EMT in the framework of this theory in two ways: through Noether procedure for translations of Minkowski time and by varying the action with respect to Minkowski metric. The problem of localizability of such an energy is also discussed. In the section \ref{en-prim} we examine the properties of the obtained EMTs in the physically interesting cases of Friedmann cosmology and the gravitational field of spherically symmetric isolated body.

\section{Energy in the embedding theory}\label{en-vlog}
\subsection{The Regge-Teitelboim gravity}\label{en-vlog-RT}
As it was mentioned in the Introduction, the approach to gravity proposed by Regge and Teitelboim \cite{regge} is based on the consideration of curved spacetime as a 4D surface locally isometrically embedded in a 10D ambient Minkowski space with one timelike direction. The sufficient number of an ambient space dimension is determined by Janet-Cartan-Friedman \cite{fridman61} theorem and can be understood intuitively by counting the degrees of freedom: 4D metric has 10 independent components. In the description of embedded surface in terms of embedding function $y^a(x^\mu)$ such a metric becomes induced and can be written using embedding function:
\begin{align}\label{metric}
g_{\mu\nu}=(\partial_\mu y^a)(\partial_\nu y^b) \eta_{ab},
\end{align}
where $\eta_{ab}$ is an ambient space metric, $a,b = 0,\ldots,9$.

It is worth noting that the embedding framework itself is a powerful tool for the studying of various geometric properties of pseudo-Riemannian manifolds. In particular, it proves useful in classification of Einstein equations solutions \cite{schmutzer} as well as in the thermodynamics of the spaces with horizon, see   \cite{statja36} and references therein. The detailed description of the formalism can be found in \cite{goenner,statja18}, so we move to the discussion of the theory of gravity in which the embedding function plays the role of dynamical variable. After the appearing of such a theory in \cite{regge}, its various forms have been repeatedly discussed \cite{deser,maia89,davkar,faddeev}, as well as the potential advantages in the construction of quantum gravity on its base.

If one takes an ordinary EH action with matter as a starting point
\disn{is1}{
S=\int d^4x\, \mathcal{L}, \qquad
\mathcal{L}=-\frac{1}{2\ka}\sqrt{-g}\,R + \mathcal{L}_\text{m}
\nom}
(where $\mathcal{L}_\text{m}$ is a matter Lagrangian density)
and substitutes \eqref{metric} into it, then after varying with respect to $y^a$ the Regge-Teitelboim (RT) equations arise, which can be written in two equivalent (if the matter EoM are satisfied) ways:
\begin{align}\label{RT}
\partial_\mu\Bigl(\sqrt{-g}(G^{\mu\nu}-\varkappa T^{\mu\nu})\partial_\nu y^a\Bigr)=0 \quad\Leftrightarrow\quad
(G^{\mu\nu}-\varkappa T^{\mu\nu})b^a_{\mu\nu}=0,
\end{align}
see details in \cite{statja18}.
Here $G^{\mu\nu}$ is an Einstein tensor and $b^a_{\mu\nu}$ is a second fundamental form of the surface:
\begin{align}\label{b_emb}
b^a_{\mu\nu}=D_\mu e^a_\nu,\qquad   e^a_\nu=\partial_\nu y^a,
\end{align}
where $D_\mu$ is a covariant derivative.

At first glance it seems that RT equations, as well as the Lagrangian $\mathcal{L}$, contain $y^a$ derivatives of more than second order, since the curvature tensor contains second-order derivatives of metric, whereas metric itself contains derivatives of $y^a$ \eqref{metric}. But this is not the case, which can be easily proven using well-known Gauss relation for curvature tensor of the surface which connects it to the second fundamental form $b^a_{\mu\nu}$ \cite{statja18}:
\begin{align}\label{gauss}
R_{\alpha \beta \mu \nu} = [ b^{e}{}_{\alpha \mu} \eta_{eg} b^{g}{}_{\beta \nu} ]_{\mu \nu}.
\end{align}
Here and hereafter we denote antisymmetrization as
 \disn{s1-26.1}{
[O_{\m\n}]_{\m\n}=O_{\m\n}-O_{\n\m}.
\nom}
As can be seen from \eqref{gauss} and \eqref{b_emb}, the curvature tensor contains derivatives of $y^a$ up to second order, so the same is true for Lagrangian  $\mathcal{L}$ and Einstein tensor, and therefore for RT equations \eqref{RT} as well.

RT equations are obviously satisfied by all solutions of Einstein equations, but the reverse is not true: RT equations possess an "extra solutions"{} for which $G^{\mu\nu}\neq\varkappa T^{\mu\nu}$. It allows to treat RT approach as modified gravity and to search for explanations of dark energy, dark matter and so on within this approach (see \cite{davids01,statja26,statja33} and references therein). This topic is beyond the scope of the present paper; instead of this we, following the authors of the original paper \cite{regge}, will treat RT approach as a search for a new set of variables for the description of gravity, which can potentially be of use in solving the problem of correct energy definition that is inherent to GR. The analogy can be drawn (see \cite{statja35} for details) with a theory of relativistic particle which is no less than 1D curved manifold embedded in 4D ambient Minkowski space. It is known that in construction of the canonical formulation in respect to particle's proper time the Hamiltonian turns out to be proportional to a constraint, whereas the energy corresponding to the proper time translations turns out to be zero.
Change of evolution parameter in the canonical formulation from proper time to ambient Minkowski time allows one to construct a non-vanishing Hamiltonian and therefore to define an energy correctly. In this paper we study the energy definitions which are based on Noether theorem, whereas various canonical formulations of embedding theory were studied in \cite{statja24,statja35,statja44}.

\subsection{Noether procedure for coordinate translations on the surface}\label{en-vlog-neter-x}
The action of embedding theory with matter is invariant with respect to translations of coordinates on the embedded surface. Let's find a pEMT corresponding to this translational invariance through Noether procedure. For the detailed description of the Noether procedure for arbitrary field theory see, e.g., \cite{Petrov-1211.3268}.

It can be easily seen that
\disn{is2}{
\dd_\al\mathcal{L}=\frac{1}{2\ka}\Bigl( 2\sqrt{-g}G^{\m\n}e_{a\m}\dd_\al\dd_\n y^a-
\dd_\n\ls \lks\sqrt{-g}g^{\m\be}\dd_\al\Gamma^\n_{\m\be}\rks^{\be\n}\rs\Bigr)+\dd_\al\mathcal{L}_\text{m}.
\nom}
We assume that matter lagrangian $\mathcal{L}_\text{m}$ depends on the fields $\varphi_A$ and their first order derivatives as well as on the metric and its first derivative.
Then we can transform $\dd_\al\mathcal{L}_\text{m}$ to the form
\disn{is2.1}{
\dd_\al\mathcal{L}_\text{m}=
\frac{\partial \mathcal{L}_{m}}{\partial\varphi_A}\partial_{\alpha}\varphi_A+
\frac{\partial \mathcal{L}_{m}}{\partial\partial_{\mu}\varphi_A}\partial_{\alpha}\dd_\m\varphi_A+
\frac{\partial \mathcal{L}_{m}}{\partial e_{\mu}^a}\dd_\al e_{\alpha}^a+
\frac{\partial \mathcal{L}_{m}}{\partial\partial_{\mu} e_{\gamma}^a}\partial_{\alpha}\dd_\m e_{\gamma}^a.
\nom}
 Using the product rule in \eqref{is2} and \eqref{is2.1} together with equations of motion \eqref{RT},  we can obtain the expression for the locally conserved
(in the sense $\dd_\m \ta^\m{}_\al$=0) pEMT:
\disn{is3}{
\ta^\n{}_\al=
\frac{\sqrt{-g}}{2\varkappa}\left(2R^{\n}_{\al}-[g^{\m\beta}\partial_{\al}\Gamma^{\n}_{\m\beta}]^{\be\n}\right)
+\hat\tau_\text{m}{}^{\n}{}_{\alpha}=\partial_{\ga}\Psi_1{}^{\ga\n}{}_{\alpha}+\hat\tau_\text{m}{}^{\n}{}_{\alpha},
\nom}
where the definition of M{\o}ller superpotential \cite{Moller}
\disn{e29.1}{
\Psi_1{}^{\ga\m}{}_\al=-\frac{1}{2\ka}\sqrt{-g}
\ls g^{\ga\be}\Ga^\m_{\be\al}-g^{\m\be}\Ga^\ga_{\be\al}\rs=
-\frac{1}{2\ka}\sqrt{-g}\ls g^{\ga\n}g^{\m\be}-g^{\m\n}g^{\ga\be}\rs\dd_\n g_{\be\al}
\nom}
was used and the matter contribution has the form
\begin{equation}
\hat\tau_\text{m}{}^{\mu}{}_{\alpha}=\frac{\partial \mathcal{L}_{m}}{\partial\partial_{\mu}\varphi_A}\partial_{\alpha}\varphi_A+\frac{\partial \mathcal{L}_{m}}{\partial e_{\mu}^a}e_{\alpha}^a+\frac{\partial \mathcal{L}_{m}}{\partial\partial_{\mu} e_{\gamma}^a}\partial_{\alpha}e_{\gamma}^a-\partial_{\gamma}\left(\frac{\partial \mathcal{L}_{m}}{\partial\partial_{\mu} e_{\gamma}^a}\right)e^a_{\alpha}-\mathcal{L}_\text{m}\delta^{\mu}_{\alpha}.
\label{Noether123}
\end{equation}

Since $\mathcal{L}_\text{m}$ depends on $e_{\mu}^a$ only through metric, one can write
\begin{align}
&\frac{\partial \mathcal{L}_\text{m}}{\partial e_{\beta}^a}=2\left(\frac{\partial \mathcal{L}_\text{m}}{\partial g_{\mu\beta}}e_{a\mu}+\frac{\partial \mathcal{L}_\text{m}}{\partial\partial_{\ga}g_{\mu\beta}}\partial_{\ga}e_{a\mu}\right),\label{h123-1}
\\
&\frac{\partial \mathcal{L}_\text{m}}{\partial\partial_{\sigma}e_{\beta}^b}=\frac{\partial \mathcal{L}_\text{m}}{\partial\partial_{\sigma}g_{\mu\beta}}e_{\mu b}+\frac{\partial \mathcal{L}_\text{m}}{\partial\partial_{\beta}g_{\mu\si}}e_{\mu b}.
\label{h123}
\end{align}
Then we substitute these expressions in \eqref{Noether123} and make use of definition of  the usual matter EMT $T^{\m\n}$ (which is on the right-hand side of Einstein equations) in the form of variational derivative of the matter action. As a result we obtain
\begin{equation}
\hat\tau_\text{m}{}^{\mu}{}_{\alpha}=-\mathcal{L}_\text{m}\delta^{\mu}_{\alpha}+\frac{\partial \mathcal{L}_{m}}{\partial\partial_{\mu}\varphi_A}\partial_{\alpha}\varphi_A+\frac{\partial \mathcal{L}_\text{m}}{\partial\partial_{\mu}g_{\beta\lambda}}\partial_{\alpha}g_{\beta\lambda}-\sqrt{-g}T^{\mu}{}_{\alpha}+\partial_{\ga}C\indices{^{\ga\mu}_{\alpha}},
\label{ds123}
\end{equation}
where
\begin{equation}
C\indices{^{\ga\mu}_{\alpha}}=\left[\frac{\partial \mathcal{L}_\text{m}}{\partial\partial_{\ga}g_{\mu\beta}}g_{\alpha\beta}\right]^{\mu\ga}.
\end{equation}
It can be shown that the sum of the first three terms in \eqref{ds123} is exactly the matter contribution $\tau_\text{m}{}^{\mu}{}_{\alpha}$ in the full Noether pEMT of GR:
\begin{equation}
\tau_\text{m}{}^{\mu}{}_{\alpha}=-\mathcal{L}_\text{m}\delta^{\mu}_{\alpha}+\frac{\partial \mathcal{L}_{m}}{\partial\partial_{\mu}\varphi_A}\partial_{\alpha}\varphi_A+\frac{\partial \mathcal{L}_\text{m}}{\partial\partial_{\mu}g_{\beta\lambda}}\partial_{\alpha}g_{\beta\lambda}.
\label{ds123-1}
\end{equation}
On the other side, it is known \cite{Mickevich1969} that $\tau_\text{m}{}^{\mu}{}_{\alpha}$ is related to $T^{\mu\nu}$ by the formula
\disn{e21}{
\ta_\text{m}{}^\m{}_\al=\sqrt{-g}\,T^\m{}_\al-\dd_\ga B^{\ga\m}{}_\al,
\nom}
where $B^{\ga\m}{}_\al=-B^{\m\ga}{}_\al$ is a certain antisymmetric expression (depending on explicit form of $S_\text{m}$) containing matter fields which are decreasing rapidly enough.
Using \eqref{e21}, we therefore obtain the final expression for $\hat\tau_\text{m}{}^{\mu}{}_{\alpha}$ (note that terms with $T^{\mu}{}_{\alpha}$ are mutually cancelled):
\begin{equation}
\hat\tau_\text{m}{}^{\mu}{}_{\alpha}=\partial_{\ga}C \indices{^{\ga\mu}_{\alpha}}-\partial_{\ga}B\indices{^{\ga\mu}_{\alpha}}.
\label{That}
\end{equation}
As a result we got an expression for the full pEMT:
\begin{equation}
\ta^{\mu}{}_{\alpha}=\partial_{\ga}\Psi\indices{_1^{\ga\mu}_{\alpha}}+\partial_{\ga}C \indices{^{\ga\mu}_{\alpha}}-\partial_{\ga}B\indices{^{\ga\mu}_{\alpha}}.
\end{equation}
Since the quantities $B^{\ga\mu}{}_{\alpha}$ and $C^{\ga\mu}{}_{\alpha}$ contain matter fields which are usually supposed to be
rapidly decreasing at the spatial infinity, the corresponding terms do not give a contribution to the conserved quantities.

As a result, the energy and momentum which are conserved according to Noether theorem due to the presence of the translation invariance with respect to the coordinates on the surface in the embedding theory, are determined by M{\o}ller superpotential (\ref{e29.1}) and completely coincide with the conserved quantities in GR.
We obtain the energy-momentum vector as an integral over the infinitely remote sphere:
\disn{d31}{
P_\al=\int\! d^3x\,\ta^0{}_\al=\int\! d^2s_i\,\Psi^{i0}{}_\al,
\nom}
which will be conserved in the assumption that
\disn{d29}{
\int\! d^2s_i\, \ta^i{}_\al=\int\! d^2s_i\, \dd_\ga\Psi^{\ga i}{}_\al=0.
\nom}
Note that it follows from \eqref{d31} that
both the energy and the momentum turn out to be zero in case of spatially closed universe.

If one replaces Lagrangian \eqref{is1} by
\disn{d43}{
{\cal L}=
-\frac{1}{2\ka} \sqrt{-g}\,
g^{\be\n}\ls \Ga^\ga_{\be\al}\Ga^\al_{\ga\n}-\Ga^\ga_{\ga\al}\Ga^\al_{\be\n}\rs+{\cal L}_\text{m},
\nom}
(in \cite{faddeev-UFN1982} one can find the discussion about advantages of using this form of action
in the asymptotically flat GR, as well as the description of the corresponding symmetry group)
then, analogously, the conserved quantities are determined by Freud superpotential
 \cite{Freud}
\disn{e85}{
\Psi_2{}^{\ga\m}{}_\al=
\frac{1}{2\ka}\frac{g_{\be\al}}{\sqrt{-g}}\dd_\n\Bigl((-g)\ls g^{\ga\n}g^{\m\be}-g^{\m\n}g^{\ga\be}\rs\Bigr),
\nom}
i. e. coincide with GR again. It is worth noting that in GR it is possible to split the full conserved pEMT $\ta^{\mu}{}_{\alpha}$ into non-conserved separately, but nevertheless non-vanishing contributions of gravity and matter, whereas in embedding theory the contribution of matter to the conserved quantities vanishes according to (\ref{That}). We also note that, in embedding theory the conserved energy, which is constructed as above, turns out to be non-localizable as in the case of GR.

\subsection{Noether procedure for ambient space coordinate translations}\label{en-vlog-neter-y}
Let us consider conservation laws that arise due to the invariance of embedding theory with respect to the translations of ambient space coordinates, under which the embedding function transforms in a following way:
\disn{is4}{
y^a(x^\m)\rightarrow y^a(x^\m)+C^a.
\nom}
This symmetry is obviously not the
spacetime
symmetry of the embedding theory, but the inner one instead.
Considering that the Lagrangian contains  derivatives of $y^a(x^\m)$ up to a second order (see section~\ref{en-vlog-RT}), the Noether procedure in this case gives the on-shell expression
\disn{is5}{
\partial_{\mu}\left(\dd_{\n}\left(\frac{\dd \mathcal{L}}{\dd\dd_{\n}e^a_{\mu}}\right)-\frac{\dd \mathcal{L}}{\dd e^a_{\mu}}\right)=0.
\nom}
Since \eqref{is1} depends on $e_{\mu}^a$ only through metric, one can use the modification of the \eqref{h123-1},\eqref{h123} (because of the fact that $\mathcal{L}$ contains second-order derivatives of metric, in contrast with $\mathcal{L}_\text{m}$ which contains only first order ones) in the equation \eqref{is5}, obtaining an expression for a locally conserved current
\disn{is6}{
j^\m_a=-2\left(\frac{\dd \mathcal{L}}{\dd g_{\mu\n}}-\dd_{\al}\frac{\dd \mathcal{L}}{\dd\dd_{\al}g_{\mu\n}}+\dd_{\al}\dd_{\be}\frac{\dd \mathcal{L}}{\dd\dd_{\alpha}\dd_{\be}g_{\mu\n}}\right)e_{\n a}=
-2e_{\n a}\frac{\de S}{\de g_{\mu\n}},\quad
\dd_{\mu} j^\m_a=0.
\nom}

For the action \eqref{is1} this current has a form
\disn{is7}{
j^{\mu}_a=-\frac{1}{\ka}\sqrt{-g}(G^{\mu\nu}-\varkappa T^{\mu\nu})e_{\nu a},
\nom}
and the condition of its local conservation coincides exactly with RT equations
Replacement of EH action by the \eqref{d43} does not change the current \eqref{is7}, whereas replacement by any other scalar density constructed from metric leads to the another expressions in the brackets on the right-hand side of \eqref{is7}, but such an expression is nevertheless vanished on-shell.

In contrast with results obtained in above sections, the locally conserved current \eqref{is7} is a tensor with respect to Lorentz transformations in ambient space and tensor density with respect to diffeomorphisms on the surface.
If one therefore interprets $j^{0}_0$ as an energy density
\disn{is7.1}{
E=\int\! d^3x\,j^{0}_0=
\int\! d^3x\,\sqrt{-g}\ls T^{0\nu}-\frac{1}{\ka}G^{0\nu}\rs\dd_\nu y_0
\nom}
(as it corresponds to ambient timelike coordinate $y^0$ translations) then such an energy will be localizable.
However, such energy vanishes for all solutions of Einstein equations, i.~e. only "extra solutions"{} of RT equations have nonzero energy
(see the remark at the end of section~\ref{en-vlog-RT}).

\section{Energy in the splitting theory}\label{en-razb}
\subsection{Gravity as a field theory in ambient spacetime}\label{en-razb-teor}
Embedding theory described in the previous section has some advantages when compared to GR from a quantization point of view, see, e. g. \cite{davkar}. However, it inherits several conceptual problems from GR, including the unavoidable usage of coordinates on a surface. To construct a coordinate-free description of gravity in \cite{statja25} was proposed a reformulation of embedding theory as a field theory in flat ambient spacetime (splitting theory). It was noticed that 4D surface $\mathcal M$ in 10D Minkowski spacetime can be defined as a constant value surface of a set of some scalar fields $z^A$ (here and hereafter $A,B,\ldots=1,\ldots,6$):
\begin{align}
z^A(y^a)=\text{const}.
\end{align}
By introduction of $z^A(y^a)$ one thus performs a "splitting"{} of Minkowski spacetime into a system of 4D surfaces. It must be noted that different functions $z^A(y^a)$ may correspond to the same splitting, if differ by transformation of "surface renumeration"{}
\begin{align}\label{gs0}
z'^A \rightarrow f^A(z^B).
\end{align}
One can introduce tensors with respect to this transformation:
\disn{gs1}{
c'^A=\frac{\dd z'^A(z)}{\dd z^B}c^B,
\nom}
the simplest of which is $\dd_a z^A$.

Using derivatives of $\dd_a z^A$ one can construct a tangential projector on $\mathcal M$ which plays an important role in the splitting theory:
\begin{align}\label{gs0.1}
\Pi_{ab}=\eta_{ab} - \po_{ab} = \eta_{ab} - \partial_a z^A \partial_b z^B w_{AB},
\end{align}
where
\begin{align}\label{w}
w^{AB}=(\partial_a z^A)(\partial_b z^B) \eta^{ab},\qquad  w_{AB}w^{BC}=\de^C_A
\end{align}
and $\po_{ab}$ is an orthogonal projector (see details in \cite{statja25}).
It is an important fact that usual derivatives of such fields as $c^A$ are not tensors in the sense of \eqref{gs1},
whereas tangent derivatives $\bar{\partial}_a \equiv \Pi^b_a \partial_b$ turn out to be tensors.
Direct calculation also gives a relation for tangential derivatives
\begin{align}\label{gs0.3}
\partial_a \left( \sqrt{|w|} \Pi^a_b A^{b\ldots c}\right)=\sqrt{|w|} \, \overline{\partial}_a (\Pi^a_b A^{b\ldots c})
\end{align}
(here $w=\det w^{AB}$), which is in some sense analogous to known relation of Riemannian geometry $\partial_\mu(\sqrt{-g} A^\mu) = \sqrt{-g} D_\mu A^\mu$.

Using projectors one can define a second fundamental form of the surface
\begin{align}\label{gs0.4}
b^a{}_{bc} = \Pi^d_b\Pi^e_c (\partial_d \Pi^f_e)\po^a_f=\Pi^d_b\Pi^e_c (\partial_d \dd_e z^A)w_{AB}\eta^{af}\dd_f z^B,
\end{align}
which allows one (through Gauss equation \eqref{gauss}) to construct Riemann and Einstein tensor as well as the scalar curvature used in the action.
It was shown in  \cite{pnan2016} that the most natural action of splitting theory is the one that proposed in \cite{statja25}
\begin{align}\label{split_EH}
S=\int d^{10} y \sqrt{|w|} \left(-\frac{1}{2\varkappa}R + \mathcal{L}_\text{m}\right),
\end{align}
where $\mathcal{L}_\text{m}$
 is a contribution of matter which in splitting theory consists of the field defined in ambient space, and it should be noted that $\mathcal{L}_\text{m}$ can depend only on tangential derivatives of these fields (see details in \cite{statja25}).
By varying \eqref{split_EH} with respect to  $z^A$ one finds that it reproduces RT equations \eqref{RT} in terms of splitting theory
\begin{align}\label{gs0.5}
(G^{bc}-\varkappa T^{bc}) b^a{}_{bc}  = 0,
\end{align}
which leads us to description of gravity in the coordinate-free formulation.

It should be noted that \eqref{split_EH} is not invariant with respect to surface renumeration \eqref{gs0}, but the corresponding equations of motion nevertheless possess such an invariance. It occurs due to the fact that after this transformation the quantity $w$  multiplies by an arbitrary function of $z^A$, which only changes the weight of the each surface $\mathcal M$ contribution in action, but since the surfaces do not interact with each other it does not affect the equations of motion.
However, for various intermediate calculations it is often useful to temporarily define coordinates on the surfaces $\mathcal M$, which can be performed in a following way. Let us define, along with Cartesian coordinates $y^a$, a set of curvilinear coordinates $\tilde{y}^a = \{x^\mu, z^A\}$ in ambient space, where $x^\mu$ are the coordinates on a surface. Then one can obtain by direct calculation of the Jacobian (see \cite{pnan2016}) that
\begin{align}\label{gs1.1}
\int d^{10} y \sqrt{|w|} (\ldots) = \int d^6 z d^4 x \sqrt{-g} (\ldots).
\end{align}
Applying this relation to \eqref{split_EH}, it is easy to notice that the action reduces to the integral over all surfaces $\mathcal M$, whereas the contribution of each surface is given by usual 4D action of gravity with matter.
It proves that the splitting theory is in some sense equivalent to the embedding theory, since although splitting theory describes many surfaces at once, there is no interaction between them and each surface has the same dynamics as in the embedding theory.

However, the action which reproduces RT (and therefore, in some sense, Einstein) equations can be chosen in
different ways. Namely, one can add the full divergence term to the
Lagrangian density
without affecting the EoM. Moreover, as it was mentioned above, the presence of such divergence terms can crucially affect the definition of energy.
These terms can also be of use when one studies the properties of variational principle usage, but such a study is beyond the scope of the present paper. Here we restrict ourselves to consideration of the action \eqref{split_EH} only, assuming that the variation of independent variables has a compact support when we derive the EoM.

\subsection{Noether procedure}\label{en-razb-neter}
Let us construct the Noether EMT with respect to the ambient space coordinate translations. The Lagrangian corresponding to action \eqref{split_EH} can be written in the form (see \cite{statja25}):
\begin{align} \label{Lagrangian}
\mathcal{L}=-\frac{1}{2 \varkappa} \sqrt{|w|} \eta^{ac} \eta^{bd}
\left [ b^e{}_{ac} \eta_{ef} b^f{}_{bd} \right]_{cd}+\sqrt{|w|}\mathcal{L}_\text{m},
\end{align}
from which, considering \eqref{gs0.1}, \eqref{w} and \eqref{gs0.4}, one can notice that this Lagrangian contains derivatives of $z^A$ up to a second order. In this case the Noether procedure gives a locally conserved current
\begin{align}
\ta^b{}_a=\frac{\partial \mathcal{L}}{\partial v^A_b}v^A_a+\frac{\partial \mathcal{L}}{\partial \partial_c v^A_b}\partial_c v^A_a-\partial_c \frac{\partial \mathcal{L}}{\partial \partial_c v^A_b}v^A_a-\mathcal{L}\delta^b_a, \label{NoetherEMT}
\end{align}
where $v^A_a=\partial_a z^A$. Using (\ref{Lagrangian}), after cumbersome calculations one can obtain EMT as a sum of gravity $t^{ba}$ and matter $\tau_\text{m}{}^{ba}$ contributions:
\begin{align}\label{Noether}
\ta^{ba}=t^{ba}+ \tau_\text{m}{}^{ba},\qquad
t^{ba}=-\frac{\sqrt{|w|}}{\varkappa}  G^{ba}.
\end{align}
Here we raised the index $a$, which is possible in splitting theory since  $\dd_b \ta^b{}_a=0$ and $\dd_b \ta^{ba}=0$ are equivalent  due to a triviality of flat metric $\eta_{ab}$.

The contribution $\tau_\text{m}{}^{ba}$ of matter fields in Noether EMT can be \cite{Mickevich1969} connected with the result of varying of $S_\text{m}$ with respect to $\eta_{ab}$, i.~e. with symmetric EMT which would be on the right-hand side of Einstein equations, if the ambient space in this approach were not flat:
\disn{gs2}{
\tau_\text{m}{}^{ba}=-2\frac{\de S_\text{m}}{\de \eta_{ba}}-\dd_c B^{cba},
\nom}
where $B^{cba}=-B^{bca}$ and the metric is chosen to be flat \textit{after} varying.
If we temporarily introduce the coordinates $x^\m$ on the surfaces, then move to the curvilinear coordinates $\tilde{y}^a = \{x^\mu, z^A\}$ in the ambient space (see the remark at the end of previous section)  and assume that in such a form (in fact, in the form of embedding theory) the Lagrangian depends on $\eta_{ba}$  only through metric $g_{\m\n}$, then
\disn{gs3}{
\frac{\de S_\text{m}}{\de \eta_{ba}(\tilde y)}=
\frac{\de S_\text{m}}{\de g_{\m\n}(\tilde y)}e_\m^b(\tilde y) e_\n^a(\tilde y)=\ns=
-\frac{\sqrt{-g(\tilde y)}}{2}T^{\m\n}(\tilde y)e_\m^b(\tilde y) e_\n^a(\tilde y)=
-\frac{\sqrt{-g(\tilde y)}}{2}T^{ba}(\tilde y),
\nom}
where we used \eqref{metric} and the fact that the variation of matter action with respect to $g_{\m\n}$ can be expressed through usual matter EMT $T^{\m\n}$.
Transforming the tensor density \eqref{gs3} back to the Minkowski ambient space coordinates $y^a$ and assuming \eqref{gs1.1} one can continue  \eqref{gs2} in the following way
\disn{gs4}{
\tau_\text{m}{}^{ba}=\sqrt{|w|}T^{ba}-\dd_c B^{cba}.
\nom}
Substituting this in \eqref{Noether}, we obtain the final expression of splitting theory EMT:
\begin{align}\label{Noether1}
\ta^{ba}=\sqrt{|w|} \left( T^{ba}-\frac{1}{\varkappa} G^{ba}\right) - \dd_c B^{cba}.
\end{align}

The quantity $B^{cba}$ depends on matter fields that usually decrease rapidly enough on the spatial directions, so the last term \eqref{Noether1} does not give a contribution (analogously to GR, see after \eqref{e21}) in the conserved energy and momentum of splitting theory, which therefore are reduced to
\begin{align}\label{gs5}
P^{a}=\int \! d^9 y \,  \ta^{0a}=\int \! d^9 y \,\sqrt{|w|}
\left( T^{0a}-\frac{1}{\varkappa}G^{0a} \right).
\end{align}
Resulting theory turns out to be the same as in section~\ref{en-vlog-neter-y}: energy vanishes for all solutions of Einstein equations, whereas non-vanishing energy corresponds only to "extra"{} solutions of RT equations (see at the end of section~\ref{en-vlog-RT}).
It is worth noting that the distribution of energy  (which density is $\sqrt{|w|}(T^{00}-G^{00}/\ka)$)  along the surface  $\mathcal{M}$  does not change under the only local transformation of the theory \eqref{gs0}, so the energy corresponding to a single surface $\mathcal{M}$ turns out to be localizable.
Such a coincidence with the results of the section~\ref{en-vlog-neter-y} can be easily explained by the fact that each of the surfaces $\mathcal{M}$ satisfies the same equations as in embedding theory, and the symmetry that was considered in both sections was translations in ambient space.

Since there is no interaction between different surfaces $\mathcal{M}$ in splitting theory and hence no energy exchange, it is interesting to pick out the contribution of single surface to the full energy. To do that one can rewrite \eqref{gs5} for $a=0$, assuming that $x^0=y^0$ and using \eqref{gs1.1} and above-mentioned curvilinear coordinates  $\tilde{y}^a = \{x^\mu, z^A\}$ in the ambient space:
\disn{gs6}{
P^0=\int \! d^6z d^3x \,\sqrt{-g}\left(T^{00}-\frac{1}{\ka} G^{00}\right).
\nom}
According to this formula one can obtain the contribution of a single surface  $\mathcal{M}$ to the full energy:
\disn{gs7}{
E_\mathcal{M}=\int\! d^3x \,\sqrt{-g}\left(T^{00}-\frac{1}{\ka} G^{00}\right).
\nom}

\subsection{Method of variation with respect to ambient space metric}\label{en-razb-metr}
As the splitting theory has the form of some field theory in flat spacetime, it is possible to use the alternative procedure of EMT construction, namely to vary the action with respect to ambient space metric (appearing result can be called "metric EMT"{} which is associated with Hilbert and Rosenfeld).

Prior to the variation one should generalize the considered field theory in the case of curved spacetime by including the interaction with nontrivial ambient space metric \textit{in a minimal way}. It means that the metric $\eta_{ab}$ becomes arbitrary and all derivatives $\partial_a$ are replaced by covariant ones $D_a$ which contain symmetric connection consisting with metric. Then one needs to vary the action with respect to metric $\eta_{ab}$, and to flatten the metric back after that. As a result one obtains a definition of \textit{a priori} symmetric EMT  $\ta^{ba}$:
\disn{gs8}{
\delta S=-\frac{1}{2} \int \! d^{10} y \sqrt{|\eta|}\, \ta^{ba} \delta \eta_{ba}.
\nom}

Note that in standard field theories in 4D Minkowski spacetime such an EMT, which corresponds to the transition to flat space in the EMT at the right-hand side of Einstein equations, can be obtained from Noether EMT by Belinfante-type procedure (see, e. g. \cite{Mickevich1969}). Moreover, in the assumption of rapid decreasing of matter fields at the spatial infinity both EMT give the same energy and momentum. The above-mentioned \emph{minimality} condition on the gravity-matter interaction in the procedure of "covariantization"{} turns out to be crucial in this case, as the addition of the contributions, which vanish in the flat limit (e. g. curvature), can alter the form of EMT as well as conserved quantities.

Let us perform a covariantization of the Lagrangian \eqref{Lagrangian} corresponding to action \eqref{split_EH}. When the ambient space becomes curved the formulas \eqref{gs0.1} and \eqref{w} remains the same, whereas \eqref{gs0.4} takes the form
\disn{gs9}{
b^a{}_{bc} = \Pi^d_b\Pi^e_c (D_d \Pi^f_e)\po^a_f,
\nom}
i. e. the only change is the replacement of $\dd_d$ by $D_d$.
There are one more change that needs to be done, namely the introduction of the multiplier $\sqrt{|\eta|}$ in the Lagrangian to provide the invariance of a volume element $d^{10}y$ in the action \eqref{split_EH}.

The simplest way to bring the covariantized expression to the form that is more convenient for the variation with respect to $\eta_{ab}$ is the usage of curvilinear ambient space coordinates $\tilde{y}^a
 = \{x^\mu, z^A\}$ mentioned above, where  $x^\m$  are auxiliary coordinates on the surfaces $\mathcal{M}$. In such curvilinear coordinates the well-known Gauss equation, which connects the curvature tensor of the surface $\mathcal{M}$ with the corresponding components of ambient space curvature tensor $R^{\text{amb}}_{abcd}$, looks very simple
\disn{gs10}{
R_{\m\n\al\be}=R^{\text{amb}}_{\m\n\al\be}+\left[ b^e{}_{\m\al} \eta_{ef} b^f{}_{\n\be}\right]_{\al\be}.
\nom}
It can be easily proven that in this coordinates
\disn{gs11}{
\Pi^{ab}=\de^a_\m\de^b_\n g^{\m\n},
\nom}
where $g^{\m\n}$ is the inverse metric of the surface $\mathcal{M}$. Using this fact one can easily obtain a corollary of \eqref{gs10}
\disn{gs12}{
R=\Pi^{ac}\Pi^{bd}R^{\text{amb}}_{abcd}+\Pi^{ac}\Pi^{bd}\left[ b^e{}_{ac} \eta_{ef} b^f{}_{bd}\right]_{cd}.
\nom}
whence, noticing that  $\Pi^{ac}b^e{}_{ad}=\eta^{ac}b^e{}_{ad}$ (it follows from \eqref{gs9} and properties of projector $\Pi^{ac}$)
one can find that
\disn{gs13}{
\eta^{ac}\eta^{bd}\left[ b^e{}_{ac} \eta_{ef} b^f{}_{bd}\right]_{cd}=R-\Pi^{ac}\Pi^{bd}R^{\text{amb}}_{abcd}.
\nom}
This relation is generally covariant, so hereafter it is possible to use any coordinate system in the ambient space besides of $\tilde{y}^a$ in which it can be obtained in the most simple way.

Using \eqref{gs13}, one can rewrite the covariantized action as a sum of contributions
\disn{gs14}{
S_1=\int \! d^{10} y \, \sqrt{|\eta|}\sqrt{|w|}\ls
-\frac{1}{2 \varkappa} R+ \mathcal{L}_\text{m}\rs
\nom}
and
\disn{gs15}{
S_2=\frac{1}{2 \varkappa} \int \! d^{10} y \, \sqrt{|\eta|} \sqrt{|w|} \Pi^{ac}\Pi^{bd}R^{\text{amb}}_{abcd}.
\nom}
It can be shown that in coordinates $\tilde{y}^a$ the contribution of $S_1$ takes the form
\disn{gs16}{
S_1=\int \! d^6z\, d^4x \, \sqrt{-g}\ls
-\frac{1}{2 \varkappa} R+ \mathcal{L}_\text{m}\rs,
\nom}
i. e. in terms of $g_{\m\n}$ only (to do that one should notice that $w=g/\tilde\eta$, where $\tilde\eta$ is the ambient space metric in coordinates $\tilde{y}^a$; the proof is the same as for \eqref{gs1.1}). It would be sensible to use this action in splitting theory if it were already formulated in \emph{curved} ambient space. However, since the original theory was formulated in \emph{flat} ambient space (like the embedding theory to which it related), the Lagrangian should be written as \eqref{Lagrangian}, which after the including of \textit{minimal} interaction with gravity leads to the resulting action $S_1+S_2$.

Let us find the variation of this action with respect to $\eta_{ab}$.
If $S_1$ is written in the form \eqref{gs16}, it depends on the quantity $\eta_{ab}$ through $g_{\m\n}$ only, the variation with respect to which is well known. Writing this variation, making use of \eqref{metric} and then rewriting the variation in arbitrary coordinates $y^a$, we find that
\disn{gs17}{
\delta S_1=\frac{1}{2 \varkappa} \int \! d^{10} y \,\sqrt{|\eta|}\sqrt{|w|} \left( G^{ba}-\varkappa T^{ba} \right)\delta \eta_{ba}.
\nom}

Now let us calculate the variation of $S_2$ \eqref{gs15}. Since to define metric EMT one should flatten the metric $\eta_{ab}$ when variation is done,  we will immediately omit all terms which are vanished in the flat limit. Considering it, we have
\disn{gs18}{
\de S_2=\frac{1}{2 \varkappa} \int \! d^{10} y \, \sqrt{|w|} \Pi^{ac}\Pi^{bd}\de R^{\text{amb}}_{abcd}.
\nom}
Noticing that in the flat limit
\begin{align}\label{gs19}
\delta R^{\text{amb}}_{abcd}=\frac{1}{2}\left[ \partial_a \partial_d \delta \eta_{bc}+ \partial_b \partial_c \delta \eta_{ad}\right]_{cd},
\end{align}
and integrating by parts, we find that
\begin{align}\label{gs20}
\delta S_2 = -\frac{1}{2 \varkappa} \int \! d^{10} y \, \partial_c \partial_d \left( \sqrt{|w|} (\Pi^{ab}\Pi^{dc}-\Pi^{ac}\Pi^{db}) \right) \delta \eta_{ba}.
\end{align}
Comparing the sum of contributions \eqref{gs17}, \eqref{gs20} with \eqref{gs8}, we obtain an expression for metric EMT of the splitting theory as a sum of gravity $t^{ba}$ and matter $\ta_{\text{m}}{}^{ba}$ contributions
\disn{gs20.1}{
\ta^{ba}=t^{ba}+ \tau_\text{m}{}^{ba},\qquad
t^{ba}=-\frac{\sqrt{|w|}}{\varkappa}  G^{ba}+\bar\ta^{ba},\qquad
\ta_{\text{m}}{}^{ba}=\sqrt{|w|}T^{ba},
\nom}
where
\disn{emt2}{
\bar\ta^{ba}=\frac{1}{\varkappa} \partial_c \partial_d \left( \sqrt{|w|} (\Pi^{ab}\Pi^{dc}-\Pi^{ac}\Pi^{db}) \right),
\nom}
so
\disn{emt2.1}{
\ta^{ba}=\sqrt{|w|}\ls T^{ba}-\frac{1}{\varkappa}G^{ba}\rs+\bar\ta^{ba}.
\nom}
Note that the same result can be obtained by the direct variation of the action corresponding to the covariantized Lagrangian \eqref{Lagrangian}, without the auxiliary coordinates $x^\m$ on the surfaces $\mathcal{M}$ and Gauss equation \eqref{gs10}, but such a calculation is a way more cumbersome.

The EMT \eqref{emt2.1} differs from the Noether one \eqref{Noether1} that obtained in section~\ref{en-razb-neter}
by a negligible (as it does not contribute in conserved quantities) term  $\dd_c B^{cba}$ and by a quantity $\bar\ta^{ba}$ \eqref{emt2}.
Noether EMT \eqref{Noether1} corresponding to Einsteinian solutions is vanished up to the negligible term mentioned above, whereas \eqref{emt2} is not, and it is easy to notice that EMT of the splitting theory calculated here is reduced exactly to \eqref{emt2} if Einstein equations are satisfied. Since $\ta^{ba}$ \eqref{gs20.1} and Noether EMT \eqref{Noether1} are locally conserved independently (as well as the term $\dd_c B^{cba}$)), the extra term \eqref{emt2} is locally conserved too: $\dd_b\bar\ta^{ba}=0$.

\subsection{The analysis of the new definition of energy and momentum}\label{en-razb-an}
In physically interesting case of Einstein solutions metric EMT of the splitting theory $\ta^{ba}$ is reduced to $\bar\ta^{ba}$. It can be written (as well as pEMT in GR) through a certain antisymmetric superpotential. To do that one should make use of \eqref{gs0.3}:
\disn{gs21}{
\bar\ta^{ba}=\partial_c \Psi^{cba},\qquad
\Psi^{cba}=\frac{1}{\varkappa}\sqrt{|w|}  \bar\partial_d (\Pi^{ab}\Pi^{dc}-\Pi^{ac}\Pi^{db}),\qquad
\Psi^{cba}=-\Psi^{bca}.
\nom}
It is interesting to discuss the localizability of such an energy. This energy, in contrast with GR one (see Introduction), is in some sense localizable as its density $\bar\ta^{00}$  cannot be set to zero by 4D coordinate transformation due to the fact that it is defined in coordinate-free formulation. However, the equations of motion (but not the action, see after \eqref{gs0.5}) of splitting theory possess the "surface renumeration"{} invariance \eqref{gs0}. It can be easily checked that $\bar\ta^{00}$  transforms inhomogeneously with respect to \eqref{gs0} and therefore can be vanished in any point, i. e. the energy is again non-localizable (even on a single surface, in contrast with Noether one, see after \eqref{gs5}) though the cause of that is not the same as in GR.

To compare the above definition of gravitational energy for Einstein solutions with known GR results we need to rewrite the expression for splitting theory energy as a sum of each surface $\mathcal{M}$ contribution. To do that, let us write the conserved energy and momentum corresponding to EMT \eqref{gs21} as a contraction:
\disn{gs22}{
n_a P^a=\int \! d^9s_b \, \bar\ta^{ba}n_a,
\nom}
where the integration is performed over an arbitrary spacelike hypersurface in ambient space. Choosing different vectors $n_a$, which are some constant (in Cartesian coordinates) functions, we obtain all conserved quantities. Let us try to write the contraction \eqref{gs22} as a sum of each surface $\mathcal{M}$ contributions. To do that, we write it in terms of $\Psi^{cba}$ using Cartesian coordinates, but in generally covariant form (using the fact that in Cartesian coordinates the standard derivative coincides with the covariant one $D_c$ and $n_a$ is a constant vector), and then we transform it to curvilinear ones $\tilde{y}^a = \{x^\mu, z^A\}$, where $x^\m$ are arbitrarily defined coordinates on the surfaces $\mathcal{M}$:
\disn{gs23}{
n_a P^a=\int \! d^9s_b \, D_c (\Psi^{cba}n_a)=
\int \! d^9 \tilde s_b \, \tilde D_c (\tilde\Psi^{cba}\tilde n_a).
\nom}
Here  $\tilde D_c$ is a covariant derivative in coordinates $\tilde y^c$ and
 $\tilde\Psi^{cba}$, $\tilde n_a$ is a result of transformations of corresponding quantities from Cartesian  $y^c$ to curvilinear coordinates $\tilde y^c$.
The integration in \eqref{gs23} is assumed to be performed over the hypersurface $\tilde y^0=const$, so this expression can be rewritten in the following form:
\disn{gs24}{
n_a P^a=
\int\! d^6 z d^3 x \,\sqrt{|\tilde\eta|}\de^0_b\tilde D_c (\tilde\Psi^{cba}\tilde n_a)=
\int\! d^6 z d^3 x \, \tilde \dd_c \ls\sqrt{|\tilde\eta|}\tilde\Psi^{c0a}\tilde n_a\rs,
\nom}
where the antisymmetry of $\tilde\Psi^{cba}$ was used together with the fact that for an arbitrary antisymmetric tensor $\tilde f^{cb}$ satisfies the relation
$\sqrt{|\tilde\eta|}\tilde D_c \tilde f^{cb}=\tilde\dd_c(\sqrt{|\tilde\eta|}\tilde f^{cb})$.
Using the antisymmetry of $\tilde\Psi^{cba}$ once more, we can rewrite the result as a sum of two terms:
\disn{gs25}{
n_a P^a=
\int\! d^6 z\, d^3 x \, \dd_i \ls\sqrt{|\tilde\eta|}\tilde\Psi^{i0a}\tilde n_a\rs+
\int\! d^6 z\, d^3 x \, \dd_A \ls\sqrt{|\tilde\eta|}\tilde\Psi^{A0a}\tilde n_a\rs,
\nom}
where indices $i$ and $A$ denote the components of  $x^\m$ and $z^A$ together constituting $\tilde y^a$, and $\dd_i\equiv\dd/\dd x^i$, $\dd_A\equiv\dd/\dd z^A$.

The second integral in \eqref{gs25} can be transformed through Gauss law into the surface integral over the infinitely remote surface in the $z^A$ space. We assume the rapid decreasing of matter fields at the spatial directions that, as it usually is in the discussion of energy and momentum in a field theory. Then at large  $z^A$ (note that all components of $z^A$ are spacelike and the only timelike coordinate in splitting theory is $x^0$) matter is absent. The surfaces $\mathcal{M}$ corresponding to these $z^A$
thus satisfy the vacuum RT equations, so one could require that they tend to planes for which $\Psi^{cba}=0$. As a result, the second term in \eqref{gs25} turns out to be zero. Note that for the surface $\mathcal{M}$, at the certain region of which the matter is present, one cannot require its flatness at the large $x^i$ (i. e. at the spacelike directions on the surface) because of the fact that the influence of matter on the surface geometry is distributed along the surfaces. For example, in case of Einsteinian solutions this influence reduces to the requirement of certain asymptotics
of metric, which restricts how rapidly $\mathcal{M}$ tends to plane at the large $x^i$. But there is no interaction between different surfaces, so equations of motion do not impose such restrictions at the large $z^A$.

The conserved energy and momentum are therefore given by the first term in \eqref{gs25} which has the form of the sum of single surface contributions. For the quantity it contains one can write
\disn{gs26}{
\sqrt{|\tilde\eta|}\tilde\Psi^{i0a}\tilde n_a=
\sqrt{|\tilde\eta|}\frac{\dd x^i}{\dd y^c}\frac{\dd x^0}{\dd y^b}\Psi^{cba}n_a=
\frac{1}{\ka}\sqrt{-g}\frac{\dd x^i}{\dd y^c}\frac{\dd x^0}{\dd y^b}\psi^{cba}n_a
\nom}
where we use \eqref{gs21} together with above-mentioned relation $w=g/\tilde\eta$ (see after \eqref{gs16}) and denote
\disn{gs27}{
\psi^{cba}=\bar\partial_d (\Pi^{ab}\Pi^{dc}-\Pi^{ac}\Pi^{db}).
\nom}
As a result, we have the following expressions for energy and momentum which correspond to EMT \eqref{gs21}:
\disn{gs28}{
P^a=
\frac{1}{\ka}\int\! d^6 z\, d^3 x \, \dd_i \ls \sqrt{-g}\,\frac{\dd x^i}{\dd y^c}\frac{\dd x^0}{\dd y^b}\psi^{cba}\rs,
\nom}
whereas the contribution of the single surface $\mathcal{M}$ has the form
\disn{gs29}{
E_\mathcal{M}=\frac{1}{\ka}\int\! d^3 x \, \dd_i \ls \sqrt{-g}\,\frac{\dd x^i}{\dd y^c}\frac{\dd x^0}{\dd y^b}\psi^{cb0}\rs=
\frac{1}{\ka}\int\! d^2s_i \sqrt{-g}\,\frac{\dd x^i}{\dd y^c}\frac{\dd x^0}{\dd y^b}\psi^{cb0},
\nom}
where the integration is performed over 2D infinitely remote spatial surface laying in $\mathcal{M}$. This is the full (gravity+matter) energy of Einstein solution in the framework of splitting theory.

\section{Embedding and splitting energy in physically interesting spacetimes}\label{en-prim}
Let us find the values of full energy of gravity+matter corresponding to its different definitions obtained above in the framework of embedding and splitting theories.

Firstly we discuss the most symmetric class of metrics, namely the cosmological model with the  FRW symmetry.
It is usually assumed in the discussion of the full energy of the system in GR that the matter is situated in some compact region, which allows to suppose the asymptotic flatness of the metric (see the detailed discussion in \cite{faddeev-UFN1982}). Therefore the cosmological case turns out to be poorly suitable for the studying of the full energy in GR:
for open and spatially flat FRW models full energy turns out to be infinite because of infinite volume of space, whereas for closed one it vanishes. The latter occurs due to the fact that the full energy in GR, with which the one that was obtained in section \ref{en-vlog-neter-x} coincides, is expressed through the integral over an infinitely remote sphere,
see ~\eqref{d31}.

However, the energy \eqref{is7.1} obtained in section \ref{en-vlog-neter-y} can be nonzero for closed universe when one considers not the Einstein solutions, but "extra solutions" of  RT equations (see end of section~\ref{en-vlog-RT}).
Let us consider "extra solutions"{} of RT equations based on the simplest embedding which has the symmetry of closed FRW model (this 5D embedding was proposed back in 1933 \cite{robertson1933})
\disn{s2-f2}{
\begin{array}{lcl}
y^0=\int dt\sqrt{\dot a(t)^2+1},                 & \qquad  &y^2=a(t)\sin\chi\,\cos\te,\\
y^1=a(t)\cos\chi,         & \qquad  &y^3=a(t)\sin\chi\,\sin\te\,\cos\ff,\\
                          & \qquad  &y^4=a(t)\sin\chi\,\sin\te\,\sin\ff,\\
\end{array}
\nom}
where dot denoted the differentiation with respect to time.
Here $a(t)$ is a time-dependent radius of curvature of the three-dimensional space
which dynamics is governed by RT equation which in this case has the following form \cite{davids01}
\disn{gs29.1}{
\dd_0\ls \rho a^3\sqrt{\dot a^2+1} - \frac{3}{\ka}a\ls\dot a^2+1\rs^\frac{3}{2}\rs=0,
\nom}
where $\rho$ is a density of matter.
It is easy to obtain that the energy \eqref{is7.1} in this case has the form
\disn{gs29.2}{
E=2\pi^2\ls \rho a^3\sqrt{\dot a^2+1} - \frac{3}{\ka}a\ls\dot a^2+1\rs^\frac{3}{2}\rs,
\nom}
i.~e. coincides (up to a numerical factor) with the constant of integration arising in the solution of RT equation \eqref{gs29.1}. It should be stressed that when the solution of this equation appears to be a solution of Einstein equations too, this constant turns out to be zero.

It can be proved that if one takes the splitting function corresponding to \eqref{s2-f2} (for the particular cases of universe expansion such a function is given in \cite{statja43}), then the energy \eqref{gs7} which is defined in the framework of splitting theory in the section \ref{en-razb-neter} is analogously reduces to the constant of integration of RT equation solution.

The way of energy defining through method used in section \ref{en-razb-metr} turns out to be poorly suitable in case of FRW symmetry. The corresponding EMT \eqref{emt2.1} contains two terms. The first term contributes to full energy like EMT \eqref{Noether1} and therefore can be written as a sum of each surface contributions, which are reduced to constants of integration of RT equation solution. The second term can be written as such only after neglecting the second integral in \eqref{gs25}, which is not possible in case of FRW symmetry, because at large $z^A$ the surfaces $\cal{M}$ cannot tend to flat ones arbitrarily fast, see the remark after \eqref{gs25}. The expression for 9D energy density $\ta^{00}$ itself in principle can be calculated for a given splitting function $z^A(y^a)$, but this quantity is not observable, because it changes under the transformations of "surface renumeration"{} \eqref{gs0}. As it was mentioned earlier (see after eq.~\eqref{gs5}), the quantity that remains unchanged after these transformations is the distribution of energy density along the surfaces $\cal{M}$, but  in the case of FRW symmetry such a quantity is not an interesting one as it reduces to a constant.
Therefore the definitions of energy obtained in the previous sections give nontrivial answers for FRW cosmology only for non-Einsteinian, "extra"{} solutions of RT equations.

Now let us consider the case of static spherically symmetric distribution of matter with mass $M$. This case seems to be more interesting for studying the problem of full energy in the presence of gravitation, as it can be assumed that the matter is situated in the compact region, and the metric is asymptotically flat. We restrict ourselves to consideration of Einsteinian solutions, when the metric is the Schwarzchild one outside of matter region,
so the only nonvanishing energy will be given by expression \eqref{gs29} obtained through method of section \ref{en-razb-metr}.

To obtain the definite value of energy one must choose the explicit form of embedding.
Among all the possible surfaces with the Schwarzchild metric (their classification for 6D ambient space can be found in \cite{statja27})
we choose the asymptotically flat embedding \cite{statja27}
\disn{gs29.3}{
\begin{array}{cl}
&\displaystyle y^0=t',\\
&y^1=r\,\cos\te,  \\
&y^2=r\,\sin\te\,\cos\ff,\\
&y^3=r\,\sin\te\,\sin\ff,\\
{\phantom{\Biggl(}}
&\displaystyle y^4=\frac{(3R)^\frac{3}{2}}{\sqrt{r}}\,\sin\ls\frac{t'}{3^{\frac{3}{2}}R}-\sqrt{\frac{R}{r}}\left( 1+\frac{r}{3R}\right)^\frac{3}{2}\rs,\\
&\displaystyle y^5=\frac{(3R)^\frac{3}{2}}{\sqrt{r}}\,\cos\ls\frac{t'}{3^{\frac{3}{2}}R}-\sqrt{\frac{R}{r}}\left( 1+\frac{r}{3R}\right)^\frac{3}{2}\rs,\\
&y^6=y^7=y^8=y^9=0\\
\end{array}
\nom}
(here $R=\ka M/4\pi$ is the Schwarzchild radius)
of the Schwarzchild metric because in this case the energy
density corresponding to \eqref{gs29} is decreasing in the spatial directions when $r\to\infty$.
The reason for this is the fact that for asymptotically flat
embeddings the projector $\Pi^{ab}$
(whose product is contained in \eqref{gs27} under the differentiation)
tends to constant at the spatial directions. Note that for all other
known embeddings of the Schwarzchild metric this condition is not
satisfied.

The corresponding to \eqref{gs29.3} splitting function can be written as $z^A(y^a)$
\begin{align}\label{gs30}
& z^1=y^4-\frac{(3R)^\frac{3}{2}}{\sqrt{r}}\sin \left( \frac{y^0}{3^{\frac{3}{2}}R}-\sqrt{\frac{R}{r}}\left( 1+\frac{r}{3R}\right)^\frac{3}{2}\right),\nonumber\\
& z^2=y^5-\frac{(3R)^\frac{3}{2}}{\sqrt{r}}\cos \left( \frac{y^0}{3^{\frac{3}{2}}R}-\sqrt{\frac{R}{r}}\left( 1+\frac{r}{3R}\right)^\frac{3}{2}\right),\\
& z^3=y^6,\quad z^4=y^7,\quad z^5=y^8,\quad z^6=y^9,\nonumber
\end{align}
where $r=\sqrt{(y^1)^2+(y^2)^2+(y^3)^2}$.
It is easy to see that each of the surfaces $z^A=const$ is a shifted
surface \eqref{gs29.3}.

The coordinates on the surfaces are chosen as $x^\m=y^\m$, i. e. they coincide with first 4 Cartesian ambient space coordinates (such a choice is allowable at least if $r$ is large enough). Then one can find that
\begin{align}\label{schwarz-psi}
\frac{\dd x^i}{\dd y^c}\frac{\dd x^0}{\dd y^b}\psi^{cb0}=
\psi^{i00}=
\ls \frac{\ka M}{2 \pi r^3}+\frac{3^3 \ka^4 M^4}{2^9\pi^4 r^6}\rs y^i=
\ls \frac{\ka M}{2 \pi r^3}+\frac{3^3 \ka^4 M^4}{2^9\pi^4 r^6}\rs x^i,
\end{align}
where $r=\sqrt{(x^1)^2+(x^2)^2+(x^3)^2}$ now.
Substituting this in \eqref{gs29} and making use of the fact that for the Schwarzchild metric in such coordinates $g=-1$, we obtain
\disn{gs31}{
E_\mathcal{M}=\int \! d^2s_i \ls \frac{M}{2 \pi r^3}+\frac{3^3 \ka^3 M^4}{2^9\pi^4 r^6}\rs x^i=2M,
\nom}
where the integration was performed over the remote sphere.

The comparison of the above result with GR one
shows that splitting theory energy coincides neither with M{\o}ller pEMT energy ($E_\mathcal{M}=M/2$),
nor with Einstein one ($E_\mathcal{M}=M$).  Since the contribution of matter in full energy is equal to $M$ in case of weak gravitational field and nonrelativistic motion of matter,
one can conclude from \eqref{gs31} that in the same approximation the gravitational energy in splitting theory is equal to $M$, whereas for M{\o}ller pEMT it is equal to $-M/2$ and for Einstein one is equal to zero.

This somewhat peculiar result is probably related to the choice of action \eqref{split_EH} which is analogous to EH one in the usual GR. The addition of certain divergence terms (see remark at the end of section~\ref{en-razb-teor}) could possibly lead to the more satisfactory value of energy, as it does in the usual GR approach, where full rest energy of isolated body calculated from first order lagrangian is equal to $M$, which coincides with special relativity.

{\bf Acknowledgements.}
The authors are grateful to A.~N.~Petrov for useful references. The work was supported by SPbU grant N~11.38.223.2015.


\end{document}